
\input phyzzx

\def\winf{$W$-infinity}
\def\win{W^+_{1+\infty}}
\def\woneplus{$W^+_{1+\infty}$}

\def\ba{{\bar A}}
\def\cd{{\cal D}}
\def\cZ{{\cal Z}}
\def\cW{{\cal W}}
\def\stwo{{\sqrt{2}}}

\def\ps{\psi}
\def\psd{\psi^{\dagger}}
\def\pat{\partial_t}
\def\pax{\partial_x}
\def\half{{1 \over 2}}
\def\wrs{W^{(r,s)}}
\def\e{\epsilon}
\def\op{[\partial_t + p \partial_q + q \partial_p]}

\def\tG{{\tilde G}}
\def\vE{{\vert E \vert}}
\def\hW{{\hat W}}
\def\sin{{\rm sin}}
\def\cos{{\rm cos}}
\def\sinh{{\rm sinh}}
\def\cosh{{\rm cosh}}
\def\coth{{\rm coth}}

\titlepage \date{December,1991.}
\hfill{IASSNS-HEP-91/79}\break
\indent\hfill{TIFR-TH-91/57}\break
\title{$W$-INFINITY WARD IDENTITIES AND CORRELATION FUNCTIONS IN THE $C=1$
MATRIX MODEL}
\author{Sumit R. Das, Avinash Dhar, Gautam Mandal\foot
{e-mail: das@tifrvax.bitnet, adhar@tifrvax.bitnet,
mandal@tifrvax.bitnet.}}
\address{ Tata Institute of Fundamental Research, Homi Bhabha Road, Bombay
400 005, India}
\andauthor{Spenta R. Wadia
\foot{Supported by DOE grant DE-FG02-90ER40542.}
\foot{e-mail: wadia@iassns.bitnet.}
\foot{On leave from Tata Institute of Fundamental Research,
Bombay 400 005, India.}
\foot{Address after January 1,1992 : Tata Institute of Fundamental Research,
Bombay 400 005, India.}}
\address{School of Natural Sciences, Institute for Advanced Study,
Princeton, NJ 08540, U.S.A}
\abstract{We explore consequences of \winf~ symmetry in the fermionic field
theory of the $c=1$ matrix model. We derive exact Ward identities relating
correlation functions of the bilocal operator. These identities can be
expressed as equations satisfied by the effective action of a {\it three}
dimensional theory and contain non-perturbative information about the
model. We use these identities to calculate the two point function of the
bilocal operator in the double scaling limit.
We extract the operator whose two point correlator has a {\it single} pole
at an (imaginary) integer value of the energy. We then rewrite the \winf~
charges in terms of operators in the matrix model and use this derive
constraints satisfied by the partition function of the matrix model with a
general time dependent potential.}
\vfill
\endpage

\Ref\DDMW{S.R. Das, A. Dhar, G. Mandal and S.R. Wadia, ETH, IAS and
Tata preprint, ETH-TH-91/30, IASSNS-HEP-91/52 and TIFR-TH-91/44, to appear
in Int. J. Mod. Phys. A.}
\Ref\DDMWA{S.R. Das, A. Dhar, G. Mandal and S.R. Wadia, IAS and
Tata preprint, IASSNS-HEP-91/72 and TIFR-TH-91/51, to appear in Mod. Phys.
Lett A.}
\Ref\GKN{D. Gross, I. Klebanov and M.J. Newmann, Nucl. Phys. B350
(1990) 621}
\Ref\MORE{G. Moore, Rutgers Preprint RU-91-12
(1991).}
\Ref\DAG{U. Danielsson and D. Gross,
Princeton Preprint PUPT-1258 (1991).}
\Ref\MOORE{G. Moore and N. Seiberg, Rutgers and Yale preprint, RU-91-29 and
YCTP-P19-91.}
\Ref\WIT{E. Witten, IAS Preprint IASSNS-HEP-91-51 (1991).}
\Ref\POLYAKOV{ I. Klebanov and A.M. Ployakov,
Princeton University preprint.}
\Ref\BERGSHOEFF{E. Bergshoeff, P.S. Howe, C.N. Pope, E. Sezgin and K.S.
Stelle, Texas A\& M, Imperial and Stony Brook preprint, CTP TAMU-25/91,
Imperial/TP/90-91/20  and  ITP-SB-91-17.}
\Ref\WIGNER{E.P. Wigner, Phys. Rev. 40(1932) 749.}
\Ref\POLCHINSKI{D. Minic, J. Polchinski and Z. Yang,
Texas preprint, UTTG-16-91.}
\Ref\GDICI{I.M. Gelfand and L.A. Dikii, Russian Math. Surveys 30 (1975)
77.}
\Ref\POLYTWO{A.M. Polyakov, Seminar at Rutgers University and private
communication.}
\Ref\KUTA{D. Kutasov, E. Martinec and N. Seiberg, Princeton and Rutgers
Preprint PUPT-1293, RU-91-49 (1991).}
\Ref\MPR{G. Moore, M. Plesser and S. Ramgoolam, Yale Preprint YCTP-P35-91.}

\chapter{Introduction.}

In recent papers [\DDMW,\DDMWA] we have developed a formulation of the
$c=1$ matrix model based on a gauge theory of fermions interacting with a
background
gauge field. This formulation was used to understand the symmetries of the
theory as well as the general Ward identities relating correlation
functions. The natural variables in the fermion theory are
{\it gauge covariant} biolocal operators,
$ \Phi (x,y,t) \equiv \ps (x,t) \psd (y,t) $,
involving fermions at separated points in eigenvalue space. The gauge
theory formulation led to a natural phase space bosonization of the model in
terms of variables belonging to  a coset space, ${\win / H}$, where $H$
denotes the group of transformations which commute with
the matrix of eigenvalues of $\Phi (x,y,t)$.

In this paper we explore in greater detail the Ward identities and their
consequences. The main result is an iteration formula for the $n$-point
correlation function of the operator $W(p,q,t)= 1/2 \int dx\; e^{ipx}\;
\psi^\dagger(x+q/2,t)\; \psi(x-q/2,t)$ which determines it in terms of all
the lower order correlation function. The one point function cannot be
determined in this fashion and needs to be calculated from the underlying
theory, as an input to this system of equations. As a consequence we also
present an exact equation for the effective action of this theory in terms
of a three dimensional field. Our equations are completely non-perturbative.
As an explicit example we calculate the two point
correlation function of the bilocal operator exactly in the double scaling
limit. From this general result we also extract the two point correlator
of the operators $W_{rs} \equiv \int dx~\psd (x,t) \{
(x + i \pax)^r, (x - i \pax)^s \} \ps $ and show that this two point
function has a single pole at energy $E = i(r-s)$. Poles at discrete
imaginary values of energy have been observed in the two point function of
matrix model operators [\GKN,\MORE,\DAG] and interpreted as
signatures of the discrete
states of the continuum $c=1$ theory coupled to gravity. However, the
correlation functions calculated in these works come as a {\it sum} of
poles, so that the corresponding operators couple to a linear combination
of these discrete states. Our result thus offers a way of isolating the
{\it specific} operator which couples to a single such state.

We then proceed to show how the general bilocal operator is related to
operators in the original matrix model by using the operator equations of
motion. This allows us to write down the Ward identities as differential
equations satisfied by the partition function considered as a functional of
a general matrix model potential, i.e. Schwinger -Dyson equations of the
matrix model itself. We expect that the
corresponding differential operators will satisfy a
$\win$ algebra. While we have not proved this in general, we have checked
by explicit computation that this is indeed true for the first four equations.
These are the "string equations" for this model and
may be regarded as generalizations of the $W_p$ constraints of the
$p$-matrix models. These string equations should enable us to understand
renormalization group flows between various multicritical points.

\chapter{Gauge Theory formulation of the $c=1$ matrix model.}

The double scaled fermion field theory is described by the action
$$ S = \int dt~dx~dy~\psd(x,t)[i\pat \delta(x-y) + \ba(x,y,t)
]\ps(y,t) \eqn\one$$
where $\ba(x,y,t) = \half(\pax^2 + x^2)~\delta(x-y)$
for the critical $c=1$ matrix model which describes the noncritical string
moving in two dimensions
Let us introduce the operator $|\Psi(t)> = \sum_n c_n |n>$, where $c_n$
are operators and $\{|n>\}$ denotes a complete basis
in the single particle Hilbert
space. The fermion field $\ps (x,t)$ is identified with the components of
this ket in the $x$-basis, $<x|\Psi(t)>$. Introducing the background gauge
field $\ba (t)$ with the components $\ba (x,y,t) \equiv <x|\ba (t) |y>$
we can write \one\ in the compact notation
$$ S = \int <\Psi(t) | i \pat + \ba (t) | \Psi (t)> \eqn\two$$
This action has the background gauge symmetry
$$\eqalign{ & | \Psi (t)> \rightarrow {\cal U} (t) | \Psi (t)> \cr
& \ba \rightarrow  {\cal U} (t) \ba {\cal U}^{\dagger} (t) + i {\cal U}
(t) \pat {\cal U}^{\dagger} (t)} \eqn\three $$
where ${\cal U} (t)$ is a unitary operator in the single particle Hilbert
space.

We now introduce the operator $W(p,q,t)$ in a "classical" phase space
$(p,q)$ by the formula
$$W(p,q,t)= 1/2 \int dx\; e^{ipx}\; \psi^\dagger(x+q/2,t)\; \psi(x-q/2,t)
\eqn\four$$
This operator may be also regarded as a generating function for the
generators $W^{(r,s)}$ of the \woneplus symmetry algebra of the critical
$c=1$ problem [\DDMW].
The $\wrs$ can be obtained from $W(p,q,t)$ by the formula
$$ \wrs = e^{-(r-s)t}~\int dp~dq~F_{rs}({p-q \over \stwo},{p+q \over
\stwo} ) W(p,q,t) \eqn\five $$
where
$$F_{rs} (a,b) = 2 \cos(\half ab)~ (i\partial_a)^r (i\partial_b)^s
\delta^{(2)} (a,b) \eqn\six$$
For example
$$ W^{(1,1)} = (\partial_q^2 - \partial_p)^2 W(p,q,t) \vert_{p=q=0}
\eqn\seven $$

The operators $W(p,q,t)$ satisfy a simple algebra

$$ [W(p,q,t), W(p',q',t)]= i\;\sin ({\hbar\over2}(pq' - qp'))
\; W(p+p', q+q',t)\eqn\fouraa$$
In \fouraa\ we have explicitly exhibited the string coupling constant
$ g_{str}\sim\hbar $, so that in the limit $\hbar \to 0$ we regain the
structure constants of the algebra of area-preserving diffeomorphisms on
the plane.
{}From \fouraa\ we can deduce the algebra of the operators $W^{(r,s)}$
defined in \five. The symmetries generated by these operators have been
previously discussed in [\DDMW] and [\MOORE] in the matrix model and by
[\WIT] and [\POLYAKOV] in the continuum formulation.
The quantum deformation of the classical algebra of area-preserving
diffeomorphisms
has been discussed in [\BERGSHOEFF].
Henceforth we shall work with $\hbar$ set to
unity unless necessary for clarity.

One can derive the equation of motion for the operator $W(p,q,t)$
using the equation of motion for the fermion field. This equation is
$$ {\rm D}_t W(p,q,t) \equiv \partial_t W(p,q,t) + \int dp'dq'~\sin \half
(pq'-qp') ~\ba (p',q',t)~W(p+p',q+q',t) = 0 \eqn\eight$$
where we have defined
$$\ba (p,q,t) \equiv {1 \over \pi} \int dx~ e^{-ipx}~
\ba(x+q/2,x-q/2,t) \eqn\nineb$$
The second term in \eight\ can be written as
$$ \int dp'dq'\int dp''dq''~C_{p,q;p',q'}^{p'',q''}
\ba(p',q',t)~W(p'',q'',t) \eqn\eightx$$
where
$$ C_{p,q;p',q'}^{p'',q''} = sin \half (pq'-p'q)~\delta(p+p'-p'') \delta
(q + q' - q'') \eqn\eighty$$
is the structure constant of the algebra \fouraa. Thus ${\rm D}_t$ is
really the $W$-infinity covariant derivative with $\ba(p,q,t)$ as the
(background) gauge field.
For the critical $c=1$ theory
$$ \ba (p,q,t) = (\partial_q^2 - \partial_p^2) \delta(p)~\delta(q)
\eqn\ninec$$
In this case \eight\ becomes
$$ \op W(p,q,t) = 0 \eqn\nined$$

Equations \four\ and \fouraa\ can be
understood in a framework that formulates
quantum mechanics in terms of a
classical phase space which is due to Wigner [\WIGNER]. To explain
this, let us introduce the Heisenberg-Weyl group with elements
$$g_{p,q}= e^{-i(q\hat P - p \hat X)} \eqn\fourab$$
where $(p,q)$ is a point in the plane ${\bf R}^2$ and $\hat P$ and $\hat
X$ satisfy the commutation relation $[\hat P, \hat X]= -i \hbar$.
The group multiplication law is given by
$$g_{p,q} g_{p',q'} = e^{i\hbar
(pq'-qp')/2} g_{p+p', q+q'} \eqn\fourea$$
Now noting that
$g_{p,q}= e^{-iq{\hat P}/2}
e^{ip{\hat X}} e^{-iq {\hat P}/2} $, we see that
$$W(p,q,t)=1/2 \int \;dx \;  \psi^\dagger(x,t)\;
g_{p,q} \; \psi(x,t) \eqn\fourd$$
One can now immediately see that
the algebra \fouraa\ is an immediate consequence of the group law
\fourea.

\chapter{Symmetries}

Among the background gauge transformations of the gauge field $\ba$
\three\ there are special ones which leave it invariant. These are
symmetries of the action \two\ generated by $\wrs$. Here we ould like to
discuss these in terms of the operators $W(p,q,t)$ introduced in \four.
Let us consider the transformations
$$\eqalign{& \delta \ps (x,t) = -\half i  \int
dp~dq~\epsilon(p,q,t)~e^{ip(x-\half q)} \ps(x-q,t) \cr
& \delta \psd (x,t) = \half i  \int
dp~dq~\epsilon(p,q,t)~e^{ip(x+\half q)} \psd (x+q,t)} \eqn\twelve$$
generated on the fermion field by $W(p,q,t)$. Equations \twelve\ express
the action of the Heisenberg-Weyl group on the fermion field.
Under these transformations
the action \two\ changes by
$$ \delta S = \int dtdpdq~ W(p,q,t) {\rm D}^{\dagger}_t \e(p,q,t)
\eqn\thirteen$$
where
$$ {\rm D}^{\dagger}_t \e(p,q,t) \equiv \partial_t \e(p,q,t) +
\int dp'dq'~\sin \half (pq'-qp') ~\ba^{*} (p',q',t)~\e (p+p',q+q',t)
\eqn\thirteena$$
and we have used the hermiticity of $A(x,y,t)$, i.e. $A^{*}(x,y,t) =
A(y,x,t)$.
Thus the symmetries of the action are the set of $\epsilon(p,q,t)$'s which
satisfy
$$ {\rm D}^{\dagger}_t \epsilon(p,q,t) = 0 \eqn\fourteen$$
For the critical $c=1$ theory $\ba^* (p,q,t) = \ba (-p,-q,t)$ is given by
\ninec. In this case \fourteen\ reads
$$ \op \e (p,q,t) = 0 \eqn\fourteenb$$
The corresponding charges are
$$ W[\e] = \int dpdq~ \e_0(p,q,t) W(p,q,t) \eqn\fifteen$$
where $\e_0(p,q,t)$ is a solution of \fourteen. It is straightforward to
check that \fourteen\ is satisfied by
$ \e (p,q,t) = e^{-(r-s)t} F_{rs}({p-q \over \stwo},{p+q \over
\stwo} ) $, thus verifying that the $\wrs$ generate the symmetry algebra
in the critical $c=1$ model.

\chapter{Ward Identities and the three dimensional effective action.}

For more general $\e (p,q,t)$, not necessarily satisfying \fourteen, the
variation of the action in \thirteen\ leads to Ward identities for the
correlation functions of $W(p,q,t)$. We now turn to a discussion of these.

Consider a general $n$-point function of the $W(p,q,t)$'s
$$ G_n (p_i,q_i,t_i) = <\mu|T( W(p_1,q_1,t_1),W(p_2,q_2,t_2) \cdots
W(p_n,q_n,t_n) )| \mu> \eqn\sixteen$$
where by $|\mu >$ we have denoted the filled fermi vacuum. In the operator
formalism the Ward identities can be derived by using the equation of motion
for $W(p,q,t)$, \eight, together with the symmetry algebra
\fouraa. For example, for the two point function we have
$$ \eqalign{{\rm D}_t G_2 (p,q,t;p',q',t') =
{\rm D}_t  \theta (t-t') <\mu| &  W(p,q,t)
W(p',q',t') |\mu> + \cr & {\rm D}_t  \theta (t'-t) <\mu| W(p',q',t')
W(p,q,t) | \mu>}$$
Using the equation of motion \eight\ and the symmetry algebra (setting
$\hbar = 1$) we get
$$\eqalign{ {\rm D}_t G_2(p,q,t;p',q',t') & =
 \delta(t-t') <\mu|[W(p,q,t),W(p',q',t')]|\mu> \cr & =
i \delta(t-t') \sin \half(pq'-p'q)~G_1 (p+p',q+q',t)}\eqn\eighteen$$

Alternatively one may use the functional integral formalism to derive the
Ward identities. In the functional formalism the $n$-point function
\sixteen\ is given by the expression
$$ G_n(p_i,q_i,t_i) = {1 \over \cZ} \int \cd \psd \cd \ps e^{iS}
\prod_{i=1}^{n} W(p_i,q_i,t_i) \eqn\nineteen$$
where $\cZ$ is the partition function. To derive the Ward identites we
make the change of variables $\ps \rightarrow \ps + \delta \ps$, $\psd
\rightarrow \psd + \delta \psd$ where the $\delta\ps$ and $\delta\psd$ are
given in \twelve. Under this change of variables, the action $S$ changes
as in \thirteen\ while the change in $W(p,q,t)$ is
$$ \delta W(p,q,t) = \int dp'dq'~ \sin \half(pq'-qp')~\e(p',q',t')
W(p+p',q+q',t) \eqn\twenty$$
Thus we get
$$\eqalign{ 0 = {1 \over \cZ} \int \cd \psd \cd \psi~e^{iS} [ i\delta
S~ & \prod_{i=1}^{n} W(p_i,q_i,t_i)~+ \cr &
\sum_{i=1}^{n} W(p_1,q_1,t_1) \cdots
\delta W(p_i,q_i,t_i) \cdots W(p_n,q_n,t_n)]}\eqn\twentya$$
Substituting \thirteen\ and \twenty\ in \twentya\ we get the Ward identity
$$ \eqalign{&{\rm D}_t G_{n+1}
(p,q,t;p_i,q_i,t_i) \cr & = i
\sum_{j=1}^{n} \delta (t-t_j)~\sin \half(pq_j
-qp_j)~G_n(p_1,q_1,t_1;\cdots; p_j+p,q_j+q,t_j;\cdots;p_n,q_n,t_n)
}\eqn\twentyone$$

Equation \twentyone\ gives a closed set of first order differential
equations for the multipoint correlation function of the $W(p,q,t)$'s. It
is appropriate to point out that this has happened because not only is the
change in $W(p,q,t)$ under \twelve\ proportional to another $W$ operator,
but more importantly the variation of the action also involves a
$W(p,q,t)$ (equation \thirteen). The situation may be compared with that
for the correlation functions of the Wilson loop operators in large-$N$
QCD where a
closed set of Dyson-Schwinger equations can be derived precisely because
the variation of both the Wilson loop operator as well as the action
involves another Wilson loop operator.
Starting with the one point function which can be calculated in the
fermion field theory \one, equation \twentyone\ can be used repeatedly to
solve for the $n$-point function.

These Ward identities can be summarized in terms of a differential
equation satisfied by the generating functional
$$ Z[J(p,q,t);\ba] \equiv \int \cd \psd \cd \psi~ {\rm exp}~\lbrack iS + \int
dpdqdt~J(p,q,t) W(p,q,t) \rbrack \eqn\newone$$
The equation is
$$ [i {\rm D}_t {\delta \over \delta J(p,q,t)} + \int dp'dq'~\sin
\half(p'q - q'p)~J(p'-p,q'-q,t)~{\delta \over \delta J(p',q',t)}] Z[J;\ba] = 0
\eqn\newtwo$$
The above equations lead to an equation satisfied by the effective action
defined by a Legendre transformation
$$ \Gamma [\Phi (p,q,t);\ba]
\equiv - {\rm log}~Z[J;\ba] + \int dp~dq~dt~ J(p,q,t)
\Phi(p,q,t) \eqn\newthree$$
Since
$$ {\delta {\rm log}~Z \over \delta J} = \Phi ~~~~~~{\delta \Gamma \over
\delta \Phi} = J \eqn\newfour$$
we have
$$ i {\rm D}_t \Phi (p,q,t) + \int dp'dq'~\sin \half
(p'q-q'p)~\Phi(p+p',q+q',t) ~{\delta \over \delta \Phi (p',q',t)}~\Gamma
[\Phi;\ba] = 0 \eqn\newfour$$
Equation \newfour\ gives an exact formulation of the $c=1$ matrix model in
terms of a three dimensional field $\Phi(p,q,t)$ \foot{A three dimensional
action for the discrete states of the $c=1$ Liouville theory has been
proposed in [\POLYAKOV].}.

For the critical $c=1$ model \twentyone\ reduces to
$$\eqalign{ & \op G_{n+1} (p_1,q_1,t;p_i,q_i,t_i) = \cr & i
\sum_{j=1}^{n} \delta (t-t_j)~\sin \half(pq_j
-qp_j)~G_n(p_1,q_1,t_1;\cdots; p_j+p,q_j+q,t_j;\cdots;p_n,q_n,t_n)}
\eqn\twentyoneb$$
Because of time translation invariance, in this case the one point
function $G_1$ is independent of time, so that we have the equation  for $G_1$
$$  (p\partial_q + q\partial_p) G_1(p,q,t) = 0 \eqn\twentytwo$$
This implies that $G_1(p,q,t)$ is a function of $(p^2-q^2)$ only, as may
be explicitly verified by a computation of $G_1$ in the critical
fermion field theory.

Before ending this general discussion on Ward identities
we would like to mention that we have so far obtained identities
satisfied by correlation functions of the generators of $W$-infinity
transformations of the theory. There are other operators in the fermion
field theory besides these, viz. operators which transform homogeneously
under $W$-infinity transformations. Examples of such operators are the
group elements of $W$-infinity. There would be Ward identities satisfied
by correlation functions of such operators as well. The situation is
analogous to WZNW models where apart from the currents there are the
primary fields $g(z,{\bar z})$ which are group elements and there are
the Knizhnik-Zamolodchikov equations involving the correlators of these
primary fields \foot{We would like to thank S. Shatashvili and E.
Verlinde for a discussion of this point.}.
\chapter{The two-point function}

In this section we focus our attention on the Ward identities for the
$c=1$ critical theory.
Of particular interest is the two point function $G_2(p,q,t;p',q',t')$ of
the $W(p,q,t)$'s since this contains information about the spectrum of the
theory. In this case, equation \twentyone\ gives
$$\op G_2(p,q,t;p',q',t') = i \delta(t-t')~ \sin \half(pq'-qp')~
G_1(p+p',q+q') \eqn\twentythree$$
where we have used the fact that $G_1$ is independent of time. In terms of
the fourier transforms
$$ W(p,q,E) \equiv \int dt~e^{-iEt}~W(p,q,t) \eqn\twentyfour$$
equation \twentythree\ can be written as
$$[iE + p \partial_q + q \partial_p] {\tilde G}_2(E;p,q;p',q') = 2 \pi
i~\sin \half (pq'-qp')~G_1 (p+p',q+q') \eqn\twentyfive$$
where we have defined ${\tilde G}_2$ by
$$ \int dt~dt'~ e^{-iEt}~e^{-iE't'}~G_2(p,q,t;p'q',t') =
\delta(E+E')~{\tilde G}_2 (E;p,q,;p',q') $$
Note that since $G_2$ is symmetric in the interchange of the arguments
$(p,q,t) \rightarrow (p',q',t')$, ${\tilde G}_2$ has the symmetry ${\tilde
G}_2 (E;p,q,;p',q') = {\tilde G}_2 (-E;p'q';p,q)$.

The zero modes of the operator $[iE + p \partial_q + q \partial_p]$ are
precisely the solutions of the equations of motion. They occur at $E_{rs}
= i(r-s)$ and are given by the functions $F_{rs}({p-q \over \stwo},{p+q \over
\stwo} )$ which were introduced in \six.

In order to solve \twentyfive\ we continue the energy $E$ to complex
values. Absorbing the $i$ in the definition of $E$ we have
$$  {\tG}_2(E;p,q;p',q') = { 1 \over
E + p \partial_q + q \partial_p}[2 \pi i~\sin \half (pq'-qp')~G_1
((p+p')^2 -(q+q')^2)] \eqn\newfive$$
Using the exponential representation of the inverse operator we get
$$ \tG_2 (E;p,q,;p',q') = \int_0^{({\rm sgn}) \times \infty}
ds~e^{-sE}~e^{-s(p\partial_q + q\partial_p)}~[2 \pi i~(\sin {Q \over
2})~G_1(R + R' + 2P)] \eqn\newsix$$
where we have defined
$$ P \equiv pp'-qq'~~~~~Q \equiv pq'-p'q~~~~~R = p^2 -q^2~~~~~R' \equiv
p'^2 -q'^2 \eqn\twentyeight$$
Now, noting that the operator $(p\partial_q + q\partial_p)$ is the
generator of Lorentz rotations in the $(p,q)$ space we can easily write
\newsix\ as
$$ \eqalign{\tG_2(E;p,q;p',q') = 2\pi i \int_0^{(sgn E) \times \infty}
ds~& e^{-sE}~\sin [\half(Q~cosh~s - P~sinh~s)] \cr & \times G_1 (R + R' +
2P~cosh~s - 2Q~sinh~s)} \eqn\twentynine$$
The solution \twentynine\ satisfies the symmetry requirement $\tG
(E;p,q,;p',q') = \tG (-E;p',q';p,q)$.
This expression for $\tG_2$ exactly matches with
the explicit calculation of the two point function for $q=q'=0$ given in
[\MORE].

In a similar way one can obtain the three point function in terms of the
two point function and the one point function and so on.

To proceed further we need to know the one point function $G_1$. This can
be computed in the fermionic theory. Using the notations and conventions
of [\MORE] \footnote*{Note that our $x$ stands for the matrix eigenvalue
while time is denoted by $t$. Also, the eigenvalue variable $\lambda$ used
in [\MORE] is $\stwo x$.} the result may be written as
$$G_1(p,q) = \half \int dx~e^{ipx} \int_{\mu}^{\infty} d\nu~\sum_{\e =
\pm} \ps_{\e}(\nu,x+\half q) \ps_{\e} (\nu,x-\half q) \eqn\thirty$$
Taking a derivative of \thirty\ with respect to $\mu$ we get
$$ \partial_\mu G_1(p,q) = -\half \int dx~e^{ipx} \sum_{\e=\pm}
\ps_{\e}(\mu,x+\half q) \ps_{\e} (\mu,x-\half q)  \eqn\thirtyone$$
Using the relation (A.12) in [\MORE] one can show that
$$\partial_\mu G_1(p,q) = {i \over 4 \stwo \pi} {\rm Re} \int_0^\infty d
\xi {e^{i\mu \xi + {i \over 4} (p^2 - q^2) \coth {\xi \over 2}} \over
\sinh {\xi \over 2}} \eqn\thirtytwo$$
The integral on the right hand side is well defined for complex $(p^2 -
q^2)$ with a small positive imaginary part. The result is then
analytically continued for real $(p^2 - q^2)$.
Now, taking a $\mu$-derivative of \twentynine\ and using \thirtytwo\ one gets
$$\eqalign{\partial_\mu \tG_2 (E;p,q;p',q') = - {1 \over 2 \stwo} &{\rm
Re} \int_0^\infty {d\xi \over \sinh{\xi \over 2}} \int_0^{({\rm sgn}~E)
\times \infty} ds~ e^{-sE} \sin \half (-P \sinh~s + Q \cosh~s) \cr &\times
{\rm exp}~ [i\mu \xi + {i \over 4} (R + R' + 2P \cosh~s - 2Q \sinh~s)
\coth {\xi \over 2}] }\eqn\thirtythreea $$
or,
$$ \eqalign{\partial_\mu \tG_2(E;p,q,;p',q')
= -{1 \over 4 \stwo} &{\rm Im} \int_0^{\infty} d \xi {e^{i\mu \xi
+ {i \over 4} (R + R') \coth {\xi \over 2}} \over \sinh {\xi \over 2}}
\cr & \times \int_0^{({\rm sgn}E) \times \infty}
ds ~e^{-sE} \lbrack {\rm exp}~(
{{iP\cosh(s - {\xi \over 2}) \over 2 \sinh {\xi \over 2}} - {iQ
\sinh(s - {\xi \over 2}) \over 2 \sinh {\xi \over 2}}}) - \cr & {\rm exp}~
({{iP\cosh(s + {\xi \over 2}) \over 2 \sinh {\xi \over 2}} - {iQ
\sinh(s + {\xi \over 2}) \over 2 \sinh {\xi \over 2}}}) \rbrack}
\eqn\thirtythreeb$$
For $q = q' = 0$ this expression for $\partial_\mu \tG_2$ matches exactly
with that given in [\MORE]. As in that reference, the integration over $s$
may be performed by changing variables from $s$ to $y = e^s$ and expanding
in inverse powers of $y$. The result is most naturally expressed in terms
of the new variables $z_{\pm},z'_{\pm}$ defined as
$$ z_{\pm} = {1 \over \stwo} (p \pm q)~~~~z'_{\pm} = {1 \over \stwo} (p'
\pm q' ) \eqn\thirtyfour$$
In terms of these variables we finally get
$$\eqalign{\partial_\mu \tG_2(& E;p,q,;p',q')  = - {1 \over 4 \stwo} {\rm
Im}~ \int_0^\infty {d\xi \over \sinh {\xi \over 2}}~
e^{i\mu \xi +{i \over 2}(z_+z_- + z'_+z'_-)\coth {\xi \over 2}} \cr &
\times \lbrack ({z_+z'_- \over z_-z'_+})^{{E \over 2}}~ 2 \pi e^{-i {\pi
\vert E \vert \over 2}}~~ {\sinh~\vE {\xi \over 2} \over \sin~\pi \vE}
{}~~J_{\vE} ({{\sqrt{z_+z_-z'_+z'_-}} \over \sinh~{\xi \over 2}})~ + \cr &
2 \sum_{n=1}^{\infty} i^n \{ ({z_+z'_- \over z_-z'_+})^{{n \over 2}}
{}~{1 \over (n-E)} + ({z'_+z_- \over z'_-z_+})^{{n \over 2}}
{}~{1 \over (n+E)} \}~ \sinh~n {\xi \over 2} ~
J_n ({{\sqrt{z_+z_-z'_+z'_-}} \over \sinh~{\xi \over 2}}) \rbrack}
\eqn\thirtyfive $$

Not surprisingly, equation \thirtyfive\ shows more structure than the
corresponding expression for $q=q'=0$. In particular, we can now obtain
the two point functions of the operators ${\hat W}_{rs} (E)$, defined in
terms of $W(p,q,E)$ (eqn. \twentyfour) by an expression similar to \five\
$$ {\hat W}_{rs} (E) = 2 (-i \partial_-)^r (-i\partial_+)^s \lbrack \cos
(z_-z_+ / 2) W(p,q,E) \rbrack_{z_{\pm} = 0} \eqn\thirtysix$$
where $\partial_{\pm} \equiv {\partial \over \partial z_{\pm}}$. The limit
$z_{\pm} \rightarrow 0$ in \thirtysix\ has to be taken carefully. In
fact, a sensible limit is defined by going to a region of the complex $E$
plane in which the expressions are well-defined.
To see how this comes about, let us compute the two point function
$<\hW_{rs} (E) W(p',q',-E)>$. We will assume that ${\rm Re}~E$
is positive and that $r < s$. The other cases can be treated
similarly. This two point function
can be projected out from \thirtyfive\ by taking suitable derivatives with
respect to $z_{\pm}$ as given in the definition of $\hW_{rs}(E)$ in
\thirtysix. We will calculate only the leading behavior for large $\mu$,
which corresponds to the planar limit. In this limit the terms in which
the derivatives in \thirtysix\ hit the cosine factor are lower order in
$\mu$. So we can set this factor to unity and directly take the
derivatives $\partial_-^r~\partial_+^s$ on $\tG_2$ in \thirtyfive. To
perform the
limit $z_{\pm} \rightarrow 0$ we first work in the region
${\rm Re}~E > (s-r)$ where the limit exists and then analytically continue
to other regions of the complex E plane
\footnote*{The limit $E \rightarrow Z$ is more
tricky and one needs to be careful while taking this limit. In fact, just
as for $q=q'=0$, the poles in the expression in square brackets in
\thirtyfive\ cancel in this limit.}.

Let us now compute the two-point function $<\hW_{rs}(E)~\hW_{r's'} (-E)>$.
As before we will assume that $E$ is positive and consider the case $r <
s$. It turns out that the above
two point function vanishes unless $(s-r) = (r'-s')$. Thus the region in
complex $E$ plane where the limit $z_\pm \rightarrow 0$ has to be
performed in order to define $\hW_{rs}(E)$ is precisely the region where
$\hW_{r's'}(-E)$ can be defined. We thus have a completely well defined
two point function. In fact, we find the leading $\mu$
behavior to be
$$ < \hW_{rs} (E) \hW_{r's'}(-E)> \sim \delta_{s-r,r'-s'}
\mu^{\half(r+s+r'+s')}~ {\rm log}~\mu~{ (s-r) \over E - (s-r)}
\eqn\thirtyseven$$

It is interesting to note that the above two point
function contains a pole at a single integer value of the energy, which is
actually a consequence of the fact that $\hW_{rs}$ are eigenoperators of
the hamiltonian. In fact a standard matrix model operator
$O_m = \int dx ~x^m \psd \ps$ may be written as binomial sum of
$\hW_{rs}$'s with
$r + s = m$. This explains why the two point function $<O_m~O_n>$ contains
a sum over poles : the operators $O_m$ are not energy
eigenoperators, but linear combinations of energy eigenoperators.

It is also interesting to note that the above expression for the two point
function is {\it exactly} reproduced in the analytic continuation of the
theory described in [\DDMW]. In the phase space formulation one
continues $t \rightarrow it$ as well as $p \rightarrow -ip$ (where $p$
denotes the single particle momentum).
Note that this analytic continuation is not the standard euclidean
continuation.
It leads to a hamiltonian $h =
-\half(p^2 +x^2)$. The fermi sea is filled from $-\infty$ to some level
$-\mu$. The operators $\hW_{rs}$ are well-defined in this framework and
their two point function may be computed in a standard manner. In the
planar limit the result is exactly \thirtyseven. However, in this
framework the "tachyon" pole corresponding to the first term in
\thirtyfive\ is not obtained since the energy levels are all discrete. The
equivalence of these two calculations is not entirely unexpected since the
calculation of two point functions of operators like $\int dx ~x^m \psd
\ps$ using harmonic oscillator wavefunctions with imaginary frequencies
[\DAG] agrees with that of [\MORE].

Finally we would like to emphasize that in the Minkowski section, $E$ in
\thirtyseven\ is purely imaginary so that in Minkowski space the poles are
at purely imaginary values of the physical energy. Consequently they
cannot be associated with states in the spectrum of the theory.

\chapter{$c=1$ Matrix Model with a general potential and String equations}

At first sight the operators $W(p,q,t)$ for $q \ne 0$ do not have an obvious
correspondence with operators of the original matrix model. This is
because the most general invariant local operator in the matrix model is
of the form ${\rm Tr}~{\rm exp}(ik M(t))$ whose expression in the fermion
field theory is simply $W(k,0,t)$. In this section we will show that the
$q \ne 0$ operators can be directly expressed as matrix model
operators through the Heisenberg equations of motion.

Consider a matrix model with an arbitrary time dependent potential $U(M(t))$.
The corresponding fermion field theory \one\ has a background gauge field
$$ \ba (x,y,t) = \half (\pax^2 - V(x,t)) \eqn\thirtyeight$$
where $V(x,t)$ denotes the corresponding scaled potential. Corresponding
to this gauge field the expression for $\ba(p,q,t)$ defined in \nineb\ is
$$ \ba(p,q,t) = (\partial_q^2 - V(i \partial_p,t)) \delta (p) ~\delta (q)
\eqn\thirtynine$$
Therefore, in this case the equation of motion \eight\ for $W(p,q,t)$ reads
$$ [\partial_t + p \partial_q - {i \over 2} \lbrack V(-i \partial_p +
\half q, t) - V(-i \partial_p - \half q, t) \rbrack ] W(p,q,t) = 0
\eqn\forty$$
This equation may be used to relate the operators
$$ W_n (p,t) \equiv \partial_q^n W(p,q,t) \vert_{q=0} \eqn\fortyone$$
to matrix model operators $O_n(t) \equiv {\rm Tr} M^n (t) = \int dx~
x^n~\psd (x,t) \psi (x,t)$. Using the above definition of $W_n (p,t)$ we
can obtain, from \forty\
$$ \pat W_n (p,t) + p W_{n+1} (p,t) - i \sum_{m=0}^n {\e_{n-m} \over
2^{n-m}} {n \choose m}~V^{(n-m)}(-i \partial_p,t) W_m(p,t) = 0,~~~~~~n =
0,1,2,\cdots \eqn\fortytwo$$
where $\e_k = 1$ for $k$ odd and zero otherwise and we have used the
notation $V^{(k)} (x,t)$ to denote $\pax^k V(x,t)$. Clearly we can use
\fortytwo\ repeatedly to relate $W_n(p,t)$ to $W_0 (p,t)$ by an equation
which contains at most $n$ derivatives with respect to time. The $W_0
(p,t)$ may be used, in turn, to generate all the operators $O_n (t)$. In fact,
$$ O_n(t) = 2 (-i \partial_p)^n W_0 (p,t) \vert_{p=0} \eqn\fortythree$$
(The factor 2 comes from the definition of $W(p,q,t$ which has a factor of
$\half$ in it).

The above relations mean that the set of operators $W_n(p,t)$ which have
no direct correspondence with the operators of the matrix model, is
equivalent to the set of operators $\pat^n W_0 (p,t)$ which are simply
$\pat^n {\rm Tr} e^{ip M(t)}$. The role of $q$ is replaced by time derivatives.

It is now clear that the equations \fortytwo\ imply an infinite set of
equations for the one point functions of the $O_n(t)$; the equation for
$O_n(t)$ involving at most $(n+1)$ time derivatives. Equivalently, these
equations may be expressed as an infinite set of constraints on the
partition function, which we may write as
$$ \cW_n(t)~\cZ = 0,~~~~~~n=0,1,2,\cdots \eqn\fortythree$$
These may be written in terms of the time dependent couplings $V_k (t)$
defined as
$$ V(x,t) = \sum_{k=1}^{\infty} V_k(t)x^k \eqn\fortythreea$$
Below we list the first few of these constraints
$$ \cW_0 (t) = \pat {\delta \over \delta V_0 (t)} \eqn\fortyfour$$
$$ \cW_1 (t) = \pat^2 {\delta \over \delta V_1 (t)} + \half
\sum_{k=1}^{\infty} k V_k(t) {\delta \over \delta V_{k-1} (t)}
\eqn\fortyfive$$
$$ \cW_2 (t) = \pat^3 {\delta \over \delta V_2 (t)} + \sum_{k=1}^{\infty}
[ (k+2) V_k(t) \pat {\delta \over \delta V_k (t)} + k \pat V_k(t) {\delta
\over \delta V_k (t)} ] \eqn\fortysix$$
$$ \eqalign{\cW_3 (t) = \pat^4 {\delta \over \delta V_3 (t)} + &{3 \over 2}
\sum_{k=1}^{\infty} [ {(k+2)(k+3) \over (k+1)} V_k(t) \pat^2 {\delta \over
\delta V_{k+1} (t)} + 3 {k(k+3) \over (k+1)} \pat V_k(t) {\delta \over
\delta V_{k+1} (t)} \cr & + {3 \over 2} k \pat^2 V_k(t) {\delta \over
\delta V_{k+1} (t)} - {3 \over 4} k(k-1)(k-2) V_k(t) {\delta \over \delta
V_{k-3} (t)} ] \cr & + {9 \over 4} \sum_{k,l=1}^{\infty} (k+l) V_k(t)
V_l(t) {\delta \over \delta V_{k+l-1} (t)} }\eqn\fortyseven$$

Alternatively, these equations may be derived by performing a set of
specific $W$ transformations with time dependent coefficients on the
fermionic fields, in such a way that the transformed action is of the same
form as the original action but with a modified potential. These
transformations are generated by the charges
$$ I_n = \sum_{k=0}^n \int dx~\psd(x,t) \e_k(x,t) [-i \pax]^k \psi (x,t)
\eqn\fortyeight$$
where the functions $\e_k(t)$ satisfy  the following set of equations
$$\eqalign{i \pat \e_m + \half \pax^2 \e_m - \half \sum_{j=1}^{n-m}
\e_{j+m} (-1)^{j+1} i^j { j+m \choose j} &\pax^j V (x,t) + i \pax
\e_{m-1} = 0 \cr & m = 1,2, \cdots (n-1)}\eqn\fortynine$$
$$ \pax \e_{n-1} + \pat \e_n = 0,~~~~~~~~\e_n = \e_n (t) $$
Clearly these equations can be used to determine any $\e_j$ in terms of
$\e_{j+1}$ and thus express all of them in terms of $\e_n$. Since
$\e_n(t)$ is a function of $t$ alone we can choose a fourier basis to
write the constraints on
on the partition function as
$$\cW_n (E) Z = [\int dx dt~e^{-iEt}
\lbrack i \pat \e_0 + \half \pax^2 \e_0 - \half
\sum_{j=1}^{n} \e_j (x,t) (-1)^{j+1} i^j \pax^j V (x,t) \rbrack {\delta \over
\delta V(x,t)}] \cZ = 0 \eqn\fifty$$

The operators $\cW$ clearly form a closed algebra. Since the equations
\fifty\ are rewritings of the original Ward identities, this algebra is
expected to be isomorphic to the $W$ - algebra \foot{We have
explicitly checked this for $n=0,1,2,3$}. The role of spin is played
by $n$, while the "mode number" is $E$. Note that the mode number can be
any real number and not necessarily an integer.

For time independent potential and for time independent
$\e_n$ in \fortyeight, $I_n$ reduce to the well known
transformations which generate the isospectral flows of the Schrodinger
operator
$$ I_n^{(0)} = (-\pax^2 + V(x))^{n \over 2} \vert_{+} \eqn\fiftyonea$$
where $\vert_{+}$ denotes the part of \fiftyonea\ which involves only
positive powers of the differential operator $-i \pax$.
and the corresponding string equations are simply given by
$$ \int dx \pax R_n (V(x)) {\delta \over \delta V(x)} \cZ = 0
\eqn\fiftytwo$$
where $R_n (V(x))$ are the Gelfand-Dickii coefficients [\GDICI].
For the special case of time independent quadratic critical
potential $V(x) = -x^2$ these isospectral deformations have been discussed
in [\POLCHINSKI]. Time dependent
transformations $I_n$ and the corresponding algebra
for the special case of $V(x) = -x^2$ have been given in
[\MOORE].
A different set of identities (once again for time independent
potentials) on the puncture operator, rather
than the partition function, have been derived in [\DAG].

The $\cW_n$
equations would be probably useful in studying the flows between various
multicritical points. In understanding such flows one must remember that
the potential $V(x,t)$ used above is a {\it scaled} potential. In general,
starting from some matrix model potential $U(M)$ in which the $N$
dependence of the various terms appear in such a way that the large-$N$
expansion is a topological expansion, the critical potential in the double
scaling limit will contain only a single monomial in $x$. To obtain a
scaled potential $V(x)$ which has various powers of $x$ the $N$ dependence
of the terms in the original matrix model potential have to be different
from the standard $N$ dependence. This does not, however, prevent us from
discussing flows between various multicritical points. Rather,
this means that when we consider, e.g.
flows between a critical point of the form $V(x) = -x^2$ and another
critical point $V(x) = - x^4$ by an interpolating potential $V(x) = ax^2 +
b x^4$, only the end points $a=-1,b=0$ and $a=0,b=-1$ correspond to
random surface theories. For generic $(a,b)$ the matrix model ${1 \over
N}$ expansion no longer correspond to a string perturbation theory.

\vskip 2.0cm

{\underbar{Note added}} : After this work was complete we learnt that I.
Klebanov and A. M. Polyakov [\POLYTWO], D. Kutasov, E. Matrinec and N.
Seiberg [\KUTA] and G. Moore, M. Plesser and S. Ramgoolam [\MPR] have also
considered the problem of constraining correlation functions using
different methods.

\refout
\end